# Photorealistic rendering of unidirectional free-space invisibility cloaks

**Jad C. Halimeh[1,*] and Martin Wegener[2]**

[1]*Physics Department and Arnold Sommerfeld Center for Theoretical Physics,
Ludwig-Maximilians-Universität München, D-80333 München, Germany*
[2]*Institut für Angewandte Physik, DFG-Center for Functional Nanostructures (CFN),
and Institut für Nanotechnologie, Karlsruhe Institute of Technology (KIT),
D-76128 Karlsruhe, Germany*
[*]*Jad.Halimeh@physik.lmu.de*

**Abstract:** Carpet or ground-plane invisibility cloaks hide an object in reflection and inhibit transmission experiments by construction. This concept has significantly reduced the otherwise demanding material requirements and has hence enabled various experimental demonstrations. In contrast, free-space invisibility cloaks should work in both reflection and transmission. The fabrication of omnidirectional three-dimensional free-space cloaks still poses significant challenges. Recently, the idea of the carpet cloak has been carried over to experiments on unidirectional free-space invisibility cloaks that only work perfectly for one particular viewing direction and, depending on the design, also for one linear polarization of light only. Here, by using photorealistic ray tracing, we visualize the performance of four types of such unidirectional cloaks in three dimensions for different viewing directions and different polarizations of light, revealing virtues and limitations of these approaches in an intuitive manner.

**OCIS codes:** (080.0080) Geometric optics; (230.3205) Invisibility cloaks; (160.3918) Metamaterials; (080.2710) Inhomogeneous optical media.

## References and links

1. J. B. Pendry, D. Schurig, and D. R. Smith, "Controlling electromagnetic fields," Science **312**, 1780-1782 (2006).
2. D. Schurig. J. B. Pendry, and D. R. Smith, "Calculation of material properties and ray tracing in transformation media," Opt. Express **14**, 9794-9804 (2006).
3. U. Leonhardt, "Optical conformal mapping," Science **312**, 1777-1780 (2006).
4. U. Leonhardt and T.G. Philbin, *Geometry and Light: The Science of Invisibility* (Dover, Mineola, 2010).
5. http://www.youtube.com/watch?v=RwgIr06OJLo
6. J. C. Halimeh and M. Wegener, "Time-of-flight imaging of invisibility cloaks," Opt. Express **20**, 63-74 (2012).
7. T. Ergin, J. Fischer, and M. Wegener, "Optical phase cloaking of 700-nm light waves in the far field by a three-dimensional carpet cloak," Phys. Rev. Lett **107**, 173901 (2011).
8. U. Leonhardt and T. G. Philbin, "General Relativity in Electrical Engineering," New J. Phys. **8**, 247 (2006).
9. J. Li and J. B. Pendry, "Hiding under the carpet: a new strategy for cloaking," Phys. Rev. Lett. **101**, 203901 (2008).
10. N. Landy and D. R. Smith, "A full-parameter unidirectional metamaterial cloak for microwaves," Nature Mater. **12**, 25-28 (2013).
11. G. von Freymann, A. Ledermann, M. Thiel, I. Staude, S. Essig, K. Busch, and M. Wegener, "Three-Dimensional Nanostructures for Photonics," Adv. Funct. Mater. **20**, 1038-1052 (2010).
12. J. Fischer and M. Wegener, "Three-dimensional optical laser lithography beyond the diffraction limit," Laser Phot. Rev. **7**, 22-44 (2013).
13. Andrew S. Glassner, *An Introduction to Ray Tracing* (Morgan Kaufmann, 1989).
14. A. Akbarzadeh and A. J. Danner, "Generalization of ray tracing in a linear inhomogeneous anisotropic medium: a coordinate-free approach," J. Opt. Soc. Am. A **27**, 2558-2562 (2010).

15. R. Schmied, J. C. Halimeh, and M. Wegener, "Conformal carpet and grating cloaks," Opt. Express **18**, 24361-24367 (2010).
16. J. C. Halimeh, R. Schmied, and M. Wegener, "Newtonian photorealistic ray tracing of grating cloaks and correlation-function-based cloaking-quality assessment," Opt. Express **19**, 6078-6092 (2011).
17. J. C. Halimeh and M. Wegener, "Photorealistic ray tracing of free-space invisibility cloaks made of uniaxial dielectrics," Opt. Express **20**, 28330-28340 (2012).
18. J. C. Halimeh, T. Ergin, J. Mueller, N. Stenger, and M. Wegener, "Photorealistic images of carpet cloaks," Opt. Express **17**, 19328-19336 (2009).
19. T. Ergin, J. C. Halimeh, N. Stenger, and M. Wegener, "Optical microscopy of 3D carpet cloaks: ray-tracing calculations," Opt. Express **18**, 20535-20545 (2010).
20. A. Greenleaf, Y. Kurylev, M. Lassas, and G. Uhlmann, "Electromagnetic Wormholes and Virtual Magnetic Monopoles from Metamaterials," Phys. Rev. Lett. **99**, 183901 (2007).
21. A. J. Danner, "Visualizing invisibility: Metamaterials-based optical devices in natural environments," Opt. Express **18**, 3332-3337 (2010).
22. J. C. Halimeh, T. Ergin, N. Stenger, and M. Wegener, "Transformationsoptik – Massgeschneiderter optischer Raum," Phys. Unserer Zeit **41**, 170-175 (2010).
23. G. Dolling, M. Wegener, S. Linden, and C. Hormann, "Photorealistic images of objects in effective negative-index materials," Opt. Express **14**, 1842–1849 (2006).

## 1. Introduction

An ideal invisibility cloak makes any macroscopic object located inside of the cloak appear to an outside observer just like empty space – for any viewing direction in three dimensions, for any polarization of the light wave, for inspection in reflection as well as in transmission, and for a broad range of carrier frequencies of light. This ideal represents a demanding benchmark of the ideas of transformation optics [1-4] that has not been accomplished experimentally so far.

However, possible more down-to-earth applications of cloaks may impose less stringent requirements. For example, consider a solar cell. One may want to reduce the shadowing effect of some opaque electric cable or mechanical support structure located above the solar cell to increase the overall energy-conversion efficiency. On a clear day at noon, sunlight approximately impinges from one direction. Thus, the desired effect can be achieved partially by surrounding the cable with a free-space invisibility cloak that works for one direction only. In this case, unlike for the full cloaking problem, image distortions would also be of minor concern – as long as the light is not blocked by the cable and eventually hits the solar cell. As a related but different example, consider an advertising column located in front of a frosted-glass bathroom window, leading to little daylight in the room. A unidirectional free-space cloak around the advertising column would increase the level of daylight in the bathroom. A pedestrian walking by the house facade could still read the advertisement on the column though. Finally, a cylindrical unidirectional free-space cloak slowly rotating around the cylinder axis and containing some fixed good to be advertised might also serve as an appealing eye catcher for shop windows: The good can be seen clearly, slowly turns into strangely distorted, disappears and one can see straight through the arrangement, the good gradually appears again, etc. In this example, minimizing image distortions in the "visible state" would obviously be quite relevant.

It should be mentioned that one can construct rather trivial polarization-independent unidirectional "invisibility cloaks" (see, *e.g.*, Ref. [5]). These devices are composed of a set of planar mirrors that guide light rays on a detour around an object to be hidden such that the direction of the emerging light rays is unchanged compared to empty space. These devices work only for one viewing direction because the auxiliary thin mirrors are only "invisible" if the incident rays lie in a plane parallel to these mirrors. However, due to the detour corresponding to the additional geometrical path length in air/vacuum, the optical path length through the apparatus becomes larger than in empty space. Such "ray cloaks" are thus distinct from true "wave cloaks", which can also properly reconstruct the optical path length, hence reconstruct the time-of-flight (TOF) [6] of a light ray or a light pulse or, equivalently, reconstruct the phase of the wave [7]. To emphasize this aspect, we will show TOF difference images below for the polarization independent cloaks. For the polarization dependent (birefringent) cloaks, the concept of a single TOF is not meaningful as a single incident pulse would generally lead to several emerging main pulses, corresponding to the different paths of ordinary and extraordinary rays.

In two dimensions, unidirectional free-space wave cloaks have already been suggested in the pioneering work of Leonhardt [3,8] in 2006. He showed by conformal coordinate transformations that a dedicated locally isotropic purely dielectric inhomogeneous material distribution is sufficient for accomplishing this goal. This implies that neither magnetism at optical frequencies nor material anisotropies are mandatory. In retrospect, intuitively speaking, this arrangement can be thought of as composed of two carpet cloaks [9,15] arranged back to back. Unfortunately, Leonhardt's particular 2006 suggestion [3,8] still contained singularities. More recently, Ref. [10] demonstrated experimentally a "perfect" [10] unidirectional free-space cloak composed of piecewise homogeneous anisotropic resonant magnetic materials at microwave frequencies in a two-dimensional waveguide geometry and for one linear polarization of the electromagnetic wave.

In this paper, we aim at visualizing the performance of unidirectional free-space cloaks in three dimensions, *i.e.*, for different viewing directions and for different linear polarizations of light. We consider four different types of designs in our visualizations based on rendering *via* ray tracing. The first two designs, strictly following Ref. [10] in terms of design and parameters, are based on piecewise homogeneous but anisotropic birefringent (magnetic) materials yielding in three dimensions two variations attributed to either using positive or negative uniaxial structures for the constituent homogeneous segments (both converge to the same material distribution in the center vertical plane. Design (i) is the positive and (ii) the negative uniaxial version. In both, birefringence inherently leads to a dependence on the direction of linear polarization. (iii) The third design is like the first two, but birefringence is eliminated by using anisotropic impedance-matched magneto-dielectric materials. In fact, the transformation that is behind this cloak allows for perfect cloaking in 3D for views along the transformation axis. Designs (i) and (ii) are merely uniaxial approximations of this cloak. (iv) The fourth design, following Refs. [3,15] in regard to using conformal mapping but not with respect to the particular transformation, uses locally isotropic (hence also polarization independent) but spatially inhomogeneous purely dielectric refractive-index distributions without singularities. Polymer structures according to design (iv) can be fabricated by three-dimensional direct-laser-writing optical lithography [11], possibly even at visible operation frequencies [12]. All designs (i)-(iv) include a metallic coating around the to-be-hidden region. Finally, virtues and limitations of all four approaches shall be discussed.

## 2. Ray-tracing approach

Ray tracing is an established technique for rendering photorealistic images of sceneries on the level of geometrical optics [2,13,14]. In our previous work [6,15-17], we have used a ray equation of motion that has been derived from Fermat's principle. We believe that this approach is particularly well suited for treating transformation-optics problems. In Ref. [16] we have discussed distributions of locally isotropic refractive indices, in Ref. [6] distributions of anisotropic yet impedance-matched magneto-dielectrics, and in Ref. [17] spatial distributions of uniaxial birefringent dielectrics. Uniaxial purely magnetic materials are analogous. Interfaces to air are discussed in Ref. [17]. This discussion can be generalized to interfaces between two birefringent materials. We refrain from repeating and extending the underlying mathematics here and rather refer the reader to this previous work, noting only that, unlike Ref. [17] where the cloak exhibited dielectric uniaxial anisotropy, here anisotropy is magnetic uniaxial and the ordinary and extraordinary refractive indices are given by $n_\text{o} = \sqrt{\epsilon\mu_\text{o}}$ and $n_\text{e} = \sqrt{\epsilon\mu_\text{e}}$, respectively, where $\mu_\text{o}$ and $\mu_\text{e}$ are the ordinary and extraordinary components of the diagonalized uniaxial permeability tensor and $\epsilon$ is the (constant scalar) permittivity of each constituent segment.

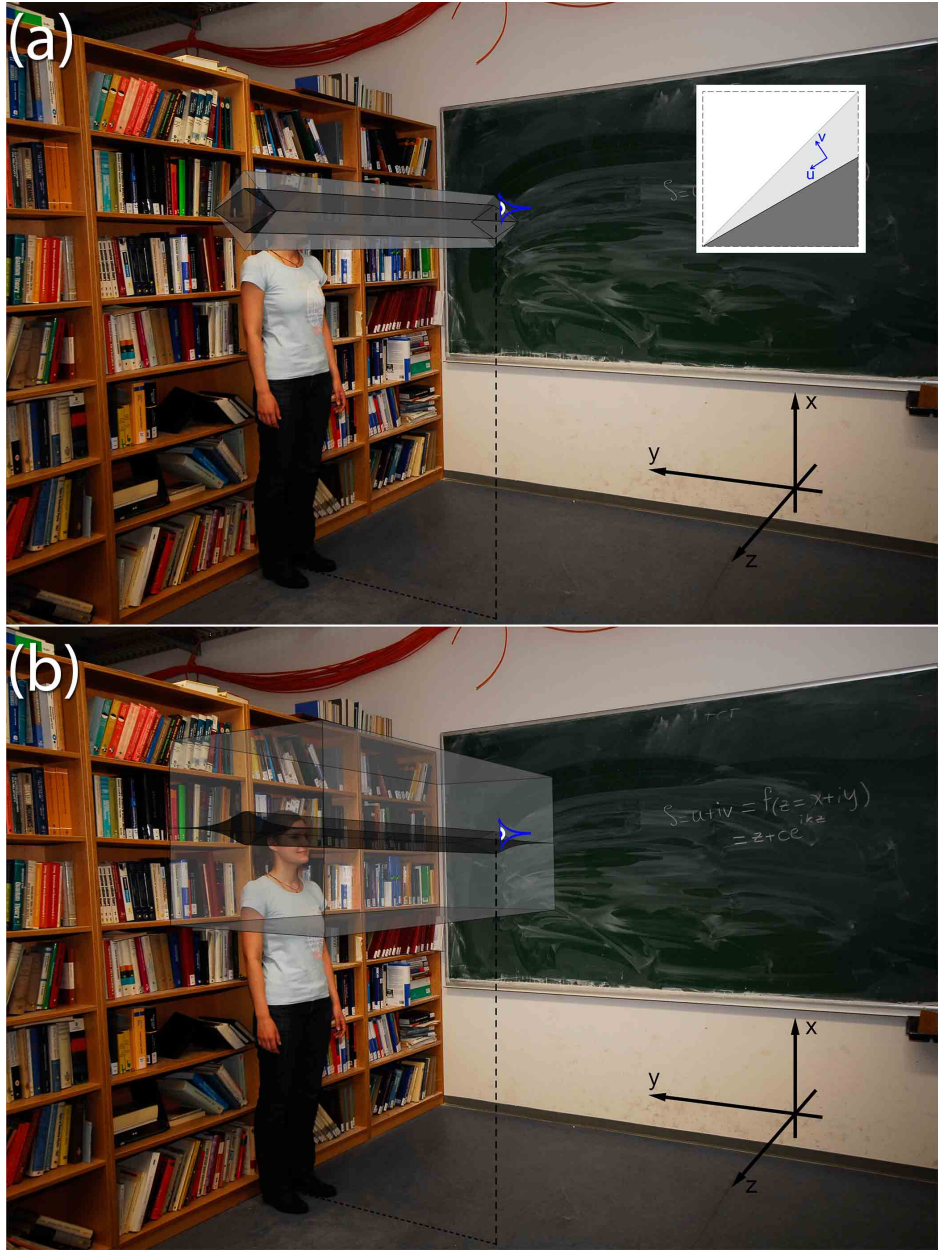

Fig. 1. Illustration showing the virtual camera (blue eye) observing either (a) the piecewise homogeneous anisotropic unidirectional free-space cloak for designs (i)-(iii) or (b) the inhomogeneous isotropic unidirectional free-space cloak for design (iv). The model stands behind the cloak. The cloak in (a) has a square cross section of side length 14.1 cm enveloping a rhombic cloaked region of side length 11.51 cm. In the inset, the local $u$-axis ($v$-axis) makes an angle of 33.54 degrees counterclockwise with respect to the $y$-axis ($x$-axis). The cloak in (b) has a rectangular cross section of width 50 cm and height 57.14 cm, enveloping a roughly Gaussian cloaked region [15]. Each cloaked region has an area of 114.83 cm$^2$. The axis of either cloak is 50 cm away from the model.

To allow for direct comparison with our previously rendered images for various other types of cloaks [6,16,17], we use the same scenery as previously. For the case of the above cloak designs (i)-(iii), the setting is illustrated in Fig. 1(a). Likewise, panel (b) depicts the setting for the cloak design (iv). A virtual point camera looks straight onto a model standing in front of a bookshelf. Curved shelf boards and/or book edges are an intuitive indicator for image distortions. Images shall be depicted for unpolarized detection as well as for detection of horizontally and vertically linear polarized light only.

## 3. Unidirectional piecewise homogeneous anisotropic square cloaks

Reference [10] considered a two-dimensional square cloak composed of a material with anisotropic magnetic permeability tensor and scalar constant electric permittivity. Here, we take exactly the same material parameters and the same square and rhombic shapes of the cloak and corresponding cloaked region, respectively, in the *xy*-plane (see Fig 1(a)). The absolute dimensions are quoted in the caption of Fig. 1. However, in three dimensions and for uniaxial constituent materials, we can either choose the magnetic uniaxial optic axis to be either along the *local u*-axis or *v*-axis leading either to positive-uniaxial (version (i): $\mu_o = 0.31$, $\mu_e = 3.27$) or negative-uniaxial (version (ii): $\mu_o = 3.27$, $\mu_e = 0.31$) segments.

(i): Results for a configuration like the one shown in Fig. 1(a) are depicted in Fig. 2. By construction in two dimensions [10], for a view along the so-called transformation axis and for horizontal polarization of light, cloaking is perfect in the center vertical plane – where horizontal polarization of light corresponds to the magnetic-induction vector lying in the vertical plane – as shown in the rendered amplitude image in Fig. 2(d). Again by construction, for the same viewing direction but for vertical polarization of light, cloaking does not work at all. Consequently, for unpolarized detection, the image is an admixture of a cloaked and an uncloaked contribution.

General viewing directions towards the left- or right-hand side of the scenery are less obvious and our renderings give a first impression on the cloaking quality. For horizontal polarization of light, the overall cloaking performance is quite good, albeit not perfect. This is due to the fact that this uniaxial approximation of cloak (iii) utilized in versions (i) and (ii) cloaks perfectly only light whose wave vector and corresponding magnetic-induction vector both lie in the vertical plane. That happens only in the center vertical plane, but upon three-dimensional views, this is no longer the case and cloaking is no longer perfect. As to be expected, vertical polarization shows no cloaking at all.

Let us now turn to deviations in direction from the unidirectional design. This means that the cloak shown in Fig. 1(a) is rotated by a certain angle $\varphi$ with respect to its own axis that is parallel to the *z*-axis. By design of the carpet cloak [9], we expect to see the projection of a flat mirror. This leads to an apparent mirror of width 19.94 rotated by $\varphi$. We only study the case of horizontal polarization as vertical polarization does not work anyway (see Fig. 2). Corresponding results are depicted in Fig. 3. Again, apart from these to-be-expected imperfections, cloaking is rather good in 3D except for significant Fresnel.

(ii): Figure 4 is like Fig. 2 but for the negative uniaxial version. Likewise, Fig. 5 is like Fig. 3 but for the negative uniaxial version. This birefringent version shows significantly better cloaking than design (i) with much weaker Fresnel reflections.

(iii): The undesired polarization dependence of designs (i) and (ii) can be eliminated by considering impedance-matched magneto-dielectric materials for which the electric permittivity tensor equals the magnetic permeability tensor. This cloak is a perfect 3D unidirectional cloak, as shown in Fig. 6. When cloaked, the corresponding mirror structure is completely invisible, and the resulting view in Fig. 6(c) is identical to the bare view (no mirror, no cloak) of Fig. 6(a). Fig. 6(d) shows the resulting TOF difference between Fig. 6(c) and Fig. 6(a). The cloaking is incredibly good, with the TOF values being limited only by machine precision. Thus, one can safely conclude that cloak (iii) is a perfect three-dimensional ray and wave unidirectional free-space cloak.

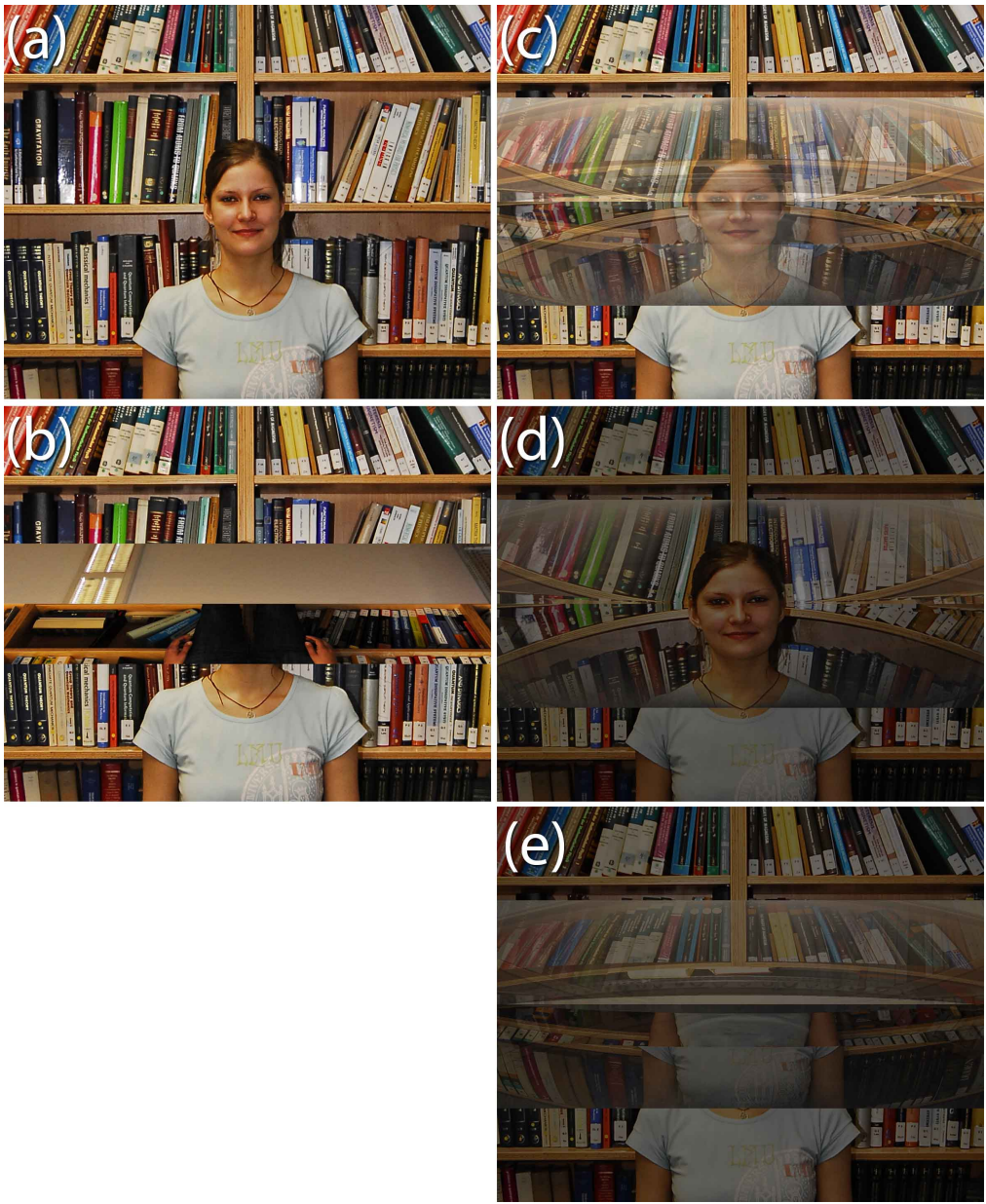

Fig. 2. The piecewise homogeneous positive uniaxial birefringent unidirectional free-space cloak under view along the transformation axis. (a) The bare field of view (FOV) as viewed by the virtual camera (see Fig. 1). (b) A rhombic mirror structure is placed in between the model and the virtual camera. (c) The cloak is placed around this structure and viewed for unpolarized light. (d) A horizontal linear polarizer is placed in front of the virtual camera, showing good cloaking behavior as extraordinary light dominates this setting. (e) A vertical linear polarizer is placed in front of the virtual camera, showing bad cloaking behavior as ordinary light is dominant for this setting.

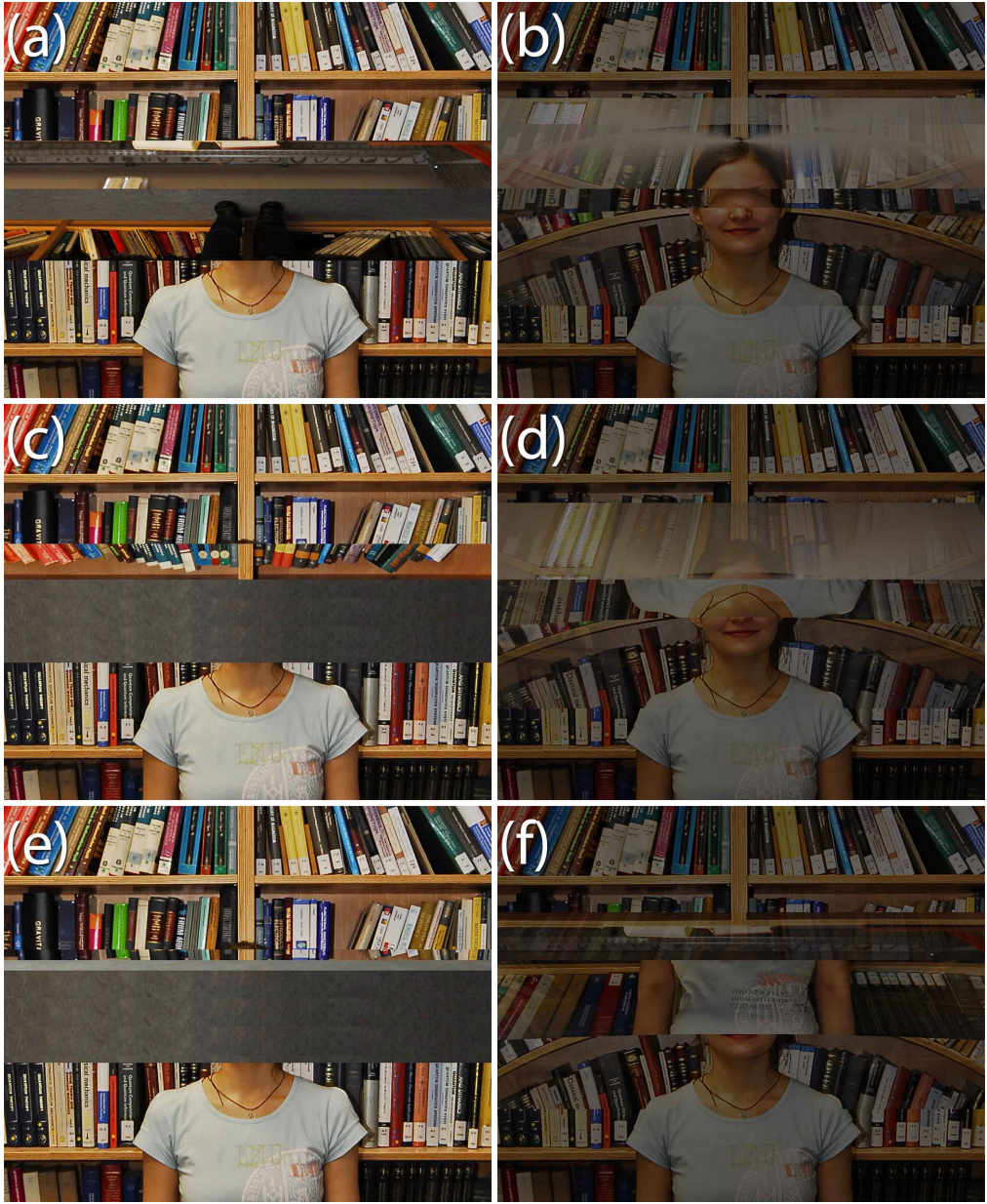

Fig. 3. The piecewise homogeneous positive uniaxial birefringent unidirectional free-space cloak under views away from the transformation axis. The bare mirror structure is shown and the corresponding cloaking result for horizontally linear polarized light, respectively for rotations of (a), (b) 5 degrees, (c), (d) 10 degrees, and (e), (f) 20 degrees around the axis of the cloak. The behavior of the unidirectional cloak for views not along the transformation axis is akin to a carpet cloak. Note the total internal reflection (TIR) in air at the first air-cloak for ordinary light interface due to $\mu_o = 0.31 \ll 1$ in the corresponding cloak segment.

Turning to deviations from views along the transformation axis, one sees that cloak (iii) is also a perfect three-dimensional carpet cloak. For any given angle $\varphi$ of rotation, cloak (iii) produces the illusion of viewing a perfectly flat mirror of width 19.94 cm rotated at the same angle. In fact, simulations (not shown) of such a mirror rotated at an angle $\varphi$ yields the exact same result shown in Fig. 7(b), (d), or (f) when $\varphi$=5, 10, or 20 degrees, respectively.

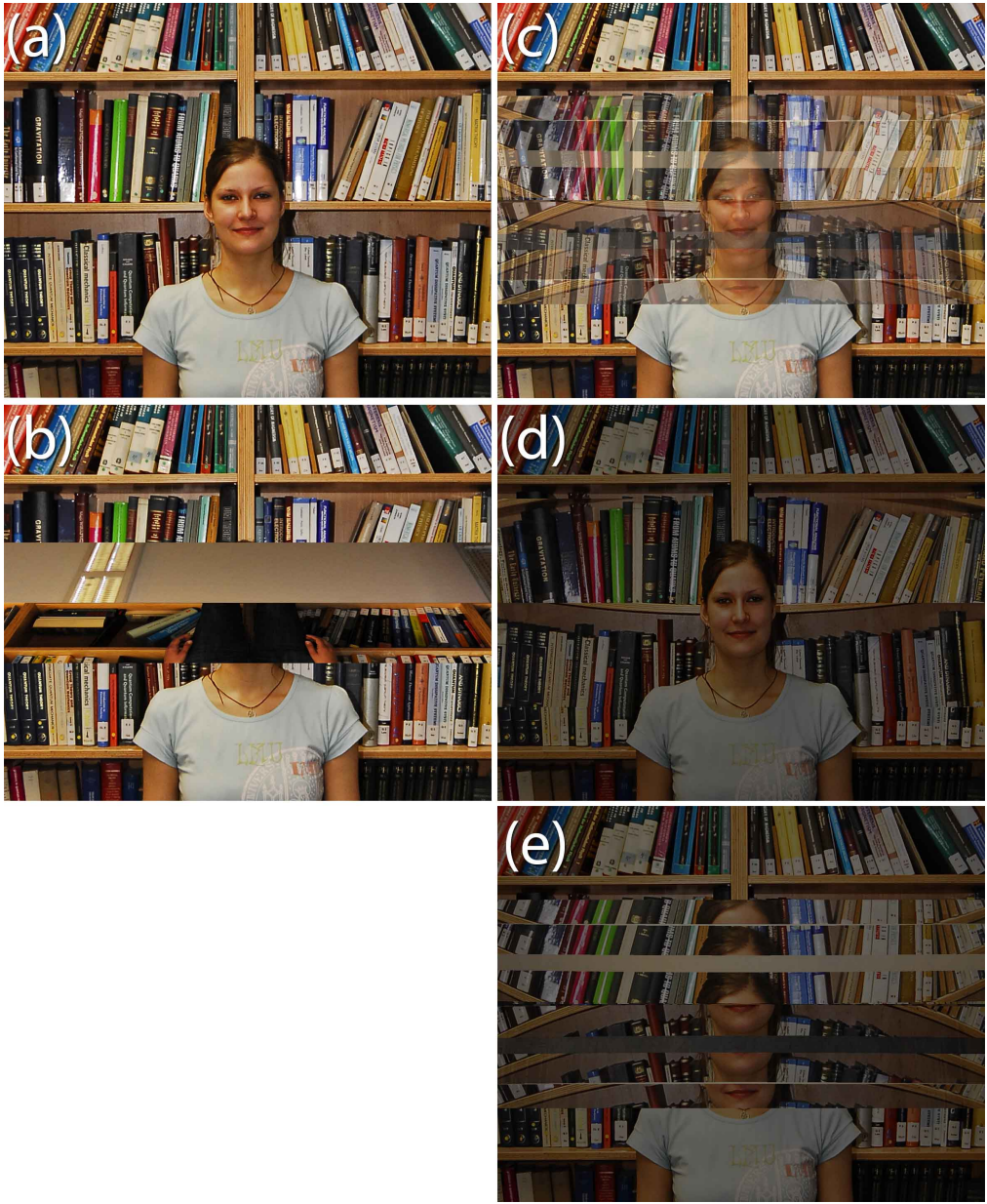

Fig. 4. Same as Fig. 2, but for a negative uniaxial variant of the piecewise homogeneous birefringent unidirectional free-space cloak. For horizontally polarized light, the cloaking behavior of this cloak is formidable as shown in (d). The step-like behavior of the vertically polarized light is due to ordinary light getting trapped in the corresponding segment of the cloak and escaping only after several total internal reflections. This is due to the fact that for this design $\mu_o = 3.27 \gg 1$ in the corresponding cloak segment. Fresnel-reflection coefficients for this cloak are of the order of less than 3%.

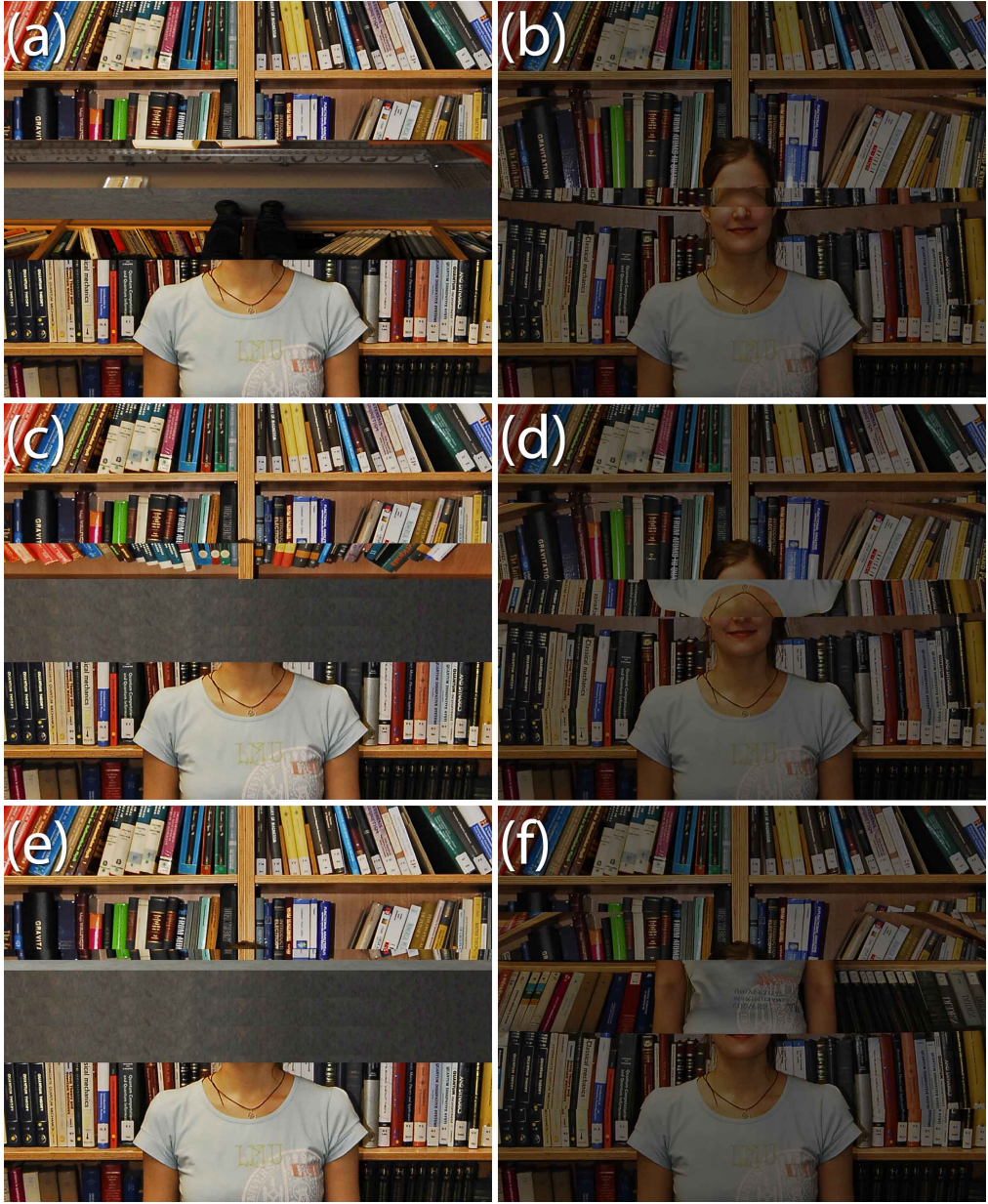

Fig. 5. Same as Fig. 3, but for a negative uniaxial variant of the piecewise homogeneous birefringent unidirectional free-space cloak. Note how here the Fresnel reflections are absent in contrast to Fig. 3. This cloak acts as a very good three-dimensional polarization dependent carpet cloak for views away from the transformation axis, leading to the illusion of a flat mirror rotated at the same angle the cloak is rotated at.

The anisotropic cloaks (i), (ii), and (iii) indeed exhibit very good to perfect unidirectional free-space cloaking. The disadvantage of design (iii), which is a perfect three-dimensional unidirectional free-space invisibility cloak, is that it requires an anisotropic tensor in its permeability and permittivity, and hence, magneto-dielectric materials, and this has proven to be quite challenging experimentally. Cloaks (i) and (ii) [10], on the other hand, require only a permeability tensor, and as Figs. 2-5 show, the unidirectional cloaking of each is quite good. In fact, the negative uniaxial version (ii) shows remarkable similarity in performance to the ideal cloak (iii), but is polarization dependent. Cloak (i) exhibits more impedance mismatch to the sides than cloak (ii). In fact, cloak (ii) shows very weak Fresnel reflections. Cloak (iii) is perfectly impedance matched to air/vacuum.

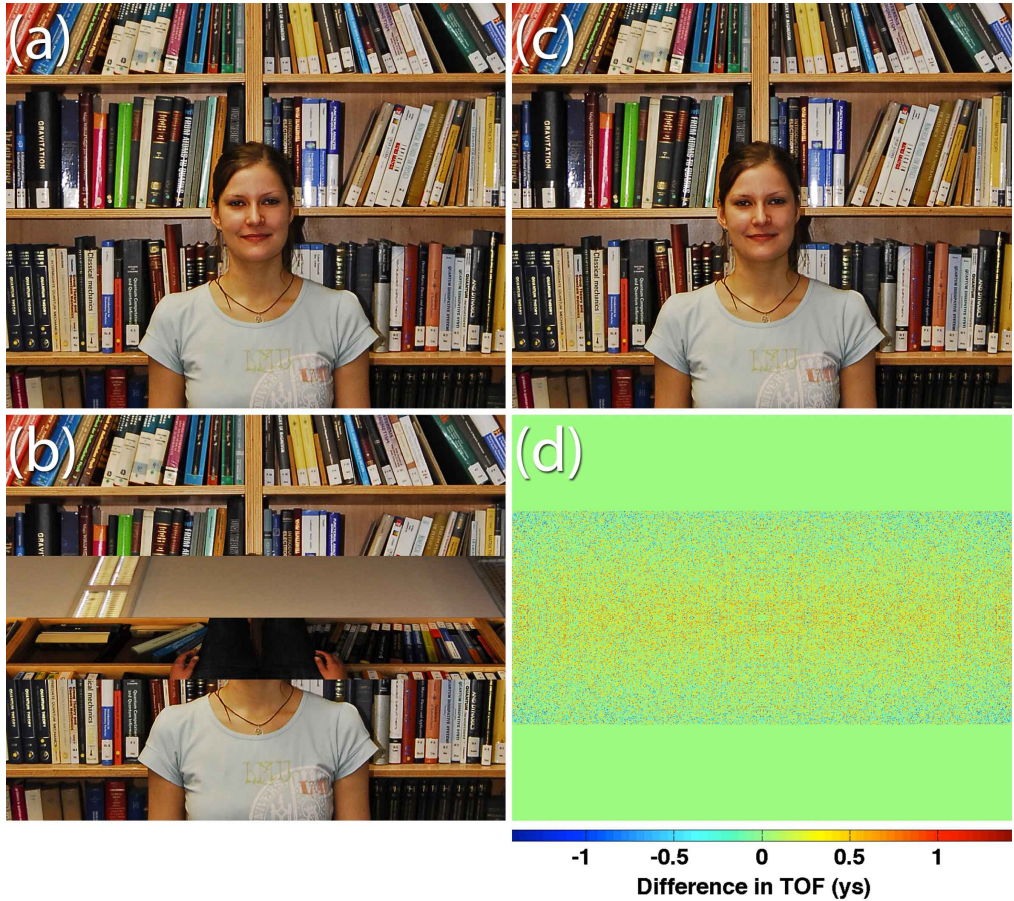

Fig. 6. Cloaking behavior of the piecewise homogeneous singly refracting unidirectional free-space cloak for a view along the transformation axis. This is a perfect cloak for this viewing direction. In (b), a rhombic mirror structure is introduced that hides the model and leads to a significant distortion of the bare FOV shown in (a). Upon introducing the singly refracting unidirectional cloak, the original bare FOV is retrieved perfectly, as shown in (c). Wave-cloaking behavior depicted in (d) by relative TOF difference is also perfect and limited only by machine precision, where the values are in the range of yoctoseconds (1 ys = $10^{-24}$ s). Note the superiority of this design to its uniaxial approximations in cloaks (i) and (ii).

It is interesting to note that each of designs (i) and (ii) show a lot of total internal reflection (TIR) behavior especially for ordinary light. Taking into consideration the first air-cloak interface, TIR occurs in air (Fresnel ordinary reflections) at that interface for design (i) since $\mu_{\text{air}} = 1 \gg \mu_o = 0.31$, while it occurs within the cloak segment at this interface for design (ii) as $\mu_o = 3.27 \gg \mu_{\text{air}} = 1$ for this design. However, for Figs. 2(d) and 4(d) that show the intended cloaking behavior for horizontally polarized light, which, even though dominated by extraordinary light still has an ordinary contribution, we see that this impedance mismatch is much more dominant in design (i) than it is in design (ii).

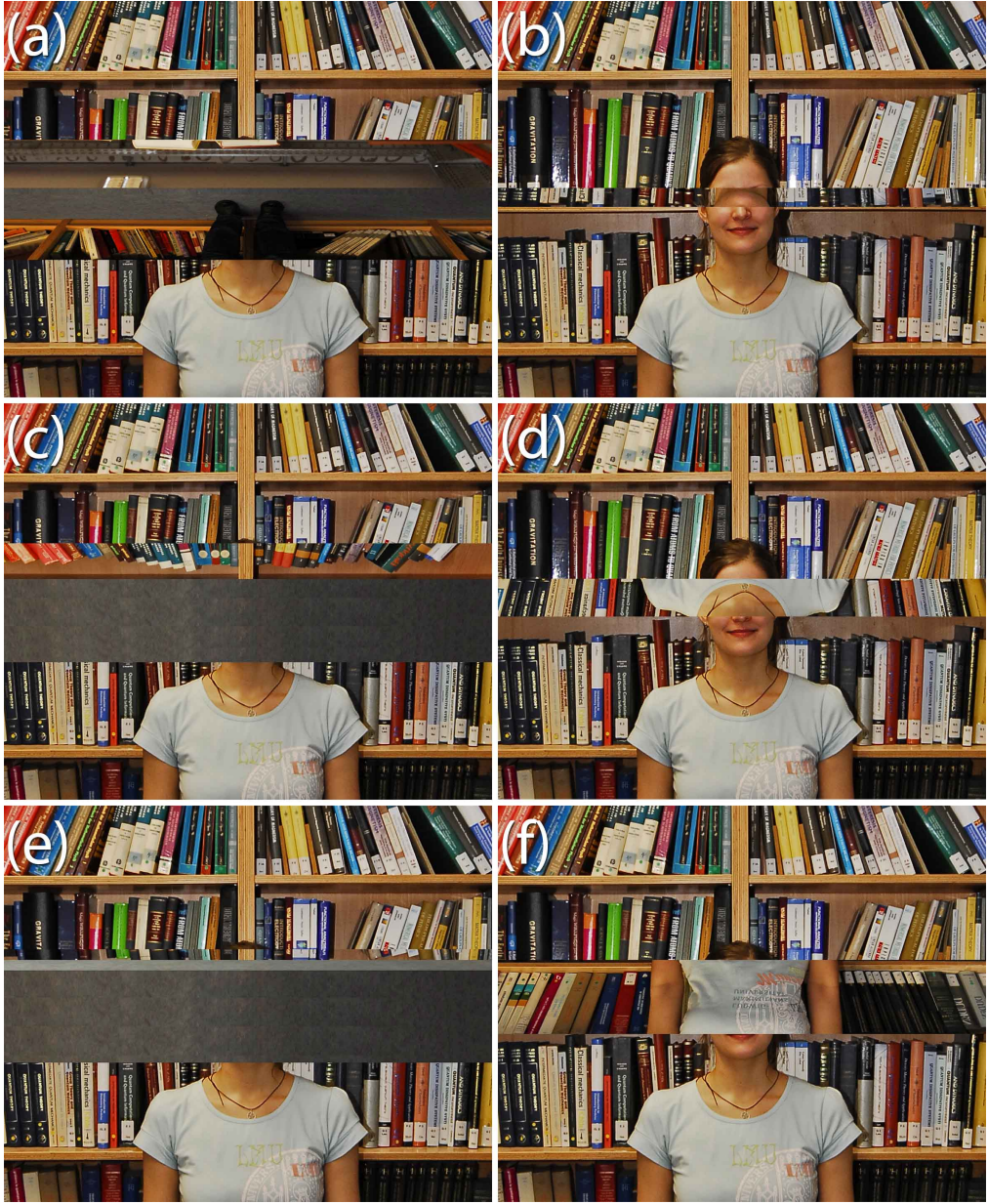

Fig. 7. Cloaking behavior of the piecewise homogeneous singly refracting unidirectional free-space cloak for views away from the transformation axis. The bare mirror structure is visualized along with its corresponding cloaked view for rotations of (a), (b) 5 degrees, (c), (d) 10 degrees, and (e), (f) 20 degrees. It is observed that this cloak acts as a perfect carpet cloak for views away from the transformation axis.

## 4. Unidirectional inhomogeneous locally isotropic double-Gaussian cloak

In Section 3, piecewise homogeneous birefringent (versions (i) and (ii)) and singly refracting (version (iii)) unidirectional free-space cloaks were investigated and their performance scrutinized. Even though version (iii) is an incredible unidirectional cloak that exhibits perfect cloaking in three dimensions, it is experimentally very difficult to realize. On the other hand, even though cloaks (i) and (ii) show remarkable cloaking in three dimensions and are experimentally feasible [10], one still faces the undesirable polarization dependence of the cloaking behavior. As such, we now turn our attention to an experimentally feasible polarization insensitive unidirectional cloak, namely the double-Gaussian unidirectional cloak. This design is based on the Gaussian carpet cloak [15,16] that is inhomogeneous and locally isotropic, and hence polarization independent. In the conformal map according to Eq. (4) of Ref. [15], we choose $w$=10.02 cm and $h$=4.25 cm. The truncating parameters are given in Fig. 1(b). This leads to a minimum (maximum) refractive index of 0.873 (1.920). Polarization independence is a very desirable feature in real-life applications, and particularly in the applications outlined in Section 1 of this manuscript. For example, in the case of solar-cell energy-conversion efficiency, cloaks (i) and (ii) would work for at most half the light impinging from the sun onto the solar panels. On the other hand, the double-Gaussian unidirectional cloak would, as will be shown and discussed shortly, transmit all of the impinging light for views along the transformation axis.

In fact, upon viewing this structure along the transformation axis, one sees very good cloaking in three dimensions. The bare view in Fig. 8(a) is severely distorted upon introducing the metal corrugation in Fig. 8(b). However, upon adding the double-Gaussian unidirectional cloak, the corrugation completely vanishes. Moreover, looking at the left- and right-hand sides of the image, one sees that there is a kind of cutting of the parts of the bare-view image around the middle horizontal line and the stretching of upper and lower parts of the image toward that line. This effect, present on the left- and right-hand sides but not the middle vertical line of the image, diminishes further away from the middle horizontal line of the image. Physically, this is due to the fact that the further from the center vertical plane the viewing direction is, the farther away light emerges from the corrugation on the other side of the cloak, rather than actually strike the corrugation itself. For an application such as solar-cell energy-conversion efficiency, for example, this is ideal, since this small cloaking imperfection in 3D is, for all purposes, perfect behavior for such an application. Also calculated and shown in Fig. 8(d) is the TOF difference between Fig. 8(c) and (a). The wave cloaking of this device is surprisingly good indicating a maximum relative error of 6%.

Fig. 9 shows views of the double-Gaussian structure for non-zero rotations around its axis, and hence, these views are not along the transformation axis. To achieve a comparable cloaked volume as its piecewise homogeneous anisotropic counterpart, the double-Gaussian unidirectional cloak exhibits a cloaked region more spread along the transformation axis. This leads the observer to see the illusion of a wider mirror upon rotations than in the case of the cloaks (i)-(iii). However, this can be avoided, in case unfavorable, by changing the values for $w$ and $h$ such that one has a thinner higher bump, and by cutting the edges in a proper manner. Such geometries are not considered here since the main interest lies in unidirectional-cloaking behavior.

Even though the conformal map that leads to the carpet cloak [15,16] utilized in the construction of the double-Gaussian unidirectional cloak is over the entire (infinite) half-plane, the actual carpet cloak itself has a finite size. This finite-size truncation leads to imperfect impedance matching. However, our calculations show that Fresnel reflections for all settings shown in Figs. 8 and 9 at the air-cloak interface are very weak, in all less than 0.2%. This is not as good as the ideal design (iii) in terms of impedance matching, but still significantly better (by orders of magnitude) than its birefringent approximations (i) and (ii). Indeed, as illustrated in Figs. 8 and 9, one sees no specular reflections on the surface of the cloak, and for all intents and purposes, for a human observer, this cloak appears to be impedance matched.

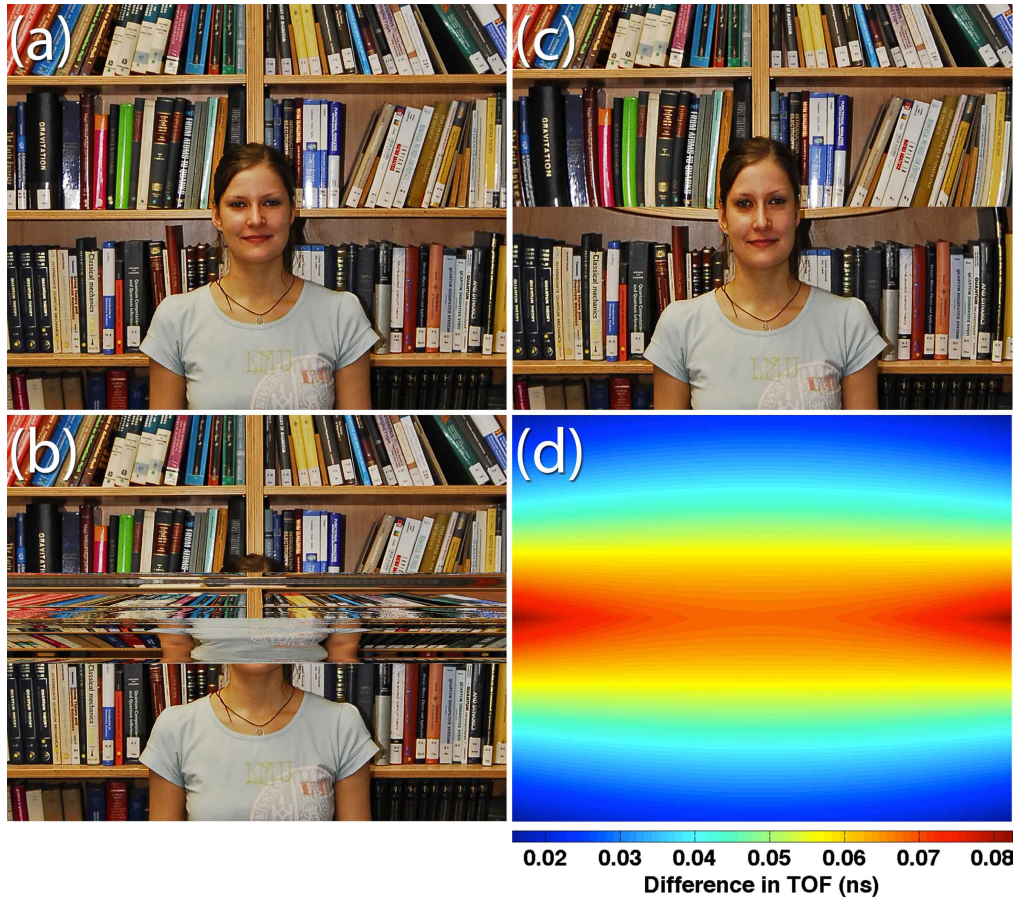

Fig. 8. The inhomogeneous locally isotropic double-Gaussian unidirectional free-space cloak for a view along the transformation axis. The Gaussian-like corrugation in (b) is cloaked amazingly well upon introduction of the cloak in (c). Worthy to note is that the cloaking performance in 3D is also very good. Moreover, the wave cloaking of this device is also formidable as shown in (d), where the TOF-difference map indicates a maximum relative error of 6% with values smaller than 0.1 nanosecond (1 ns = $10^{-9}$ s). Fresnel-reflection coefficients for this setting are of the order of 0.01% or less.

## 5. Conclusion

We have rendered photorealistic images of sceneries in three dimensions including unidirectional free-space invisibility cloaks. Compared to omnidirectional free-space cloaks, the material requirements can be relaxed for the unidirectional case. The birefringent designs (i) and (ii) work well for one particular viewing direction (along the transformation axis), but the performance deteriorates for angles departing from this view. Furthermore, due to the use of birefringence, the devices clearly work for one linear polarization of light only. This aspect is eliminated for the corresponding anisotropic but impedance-matched design (iii), which uses magneto-dielectric materials – at the expense of being much more demanding to realize.

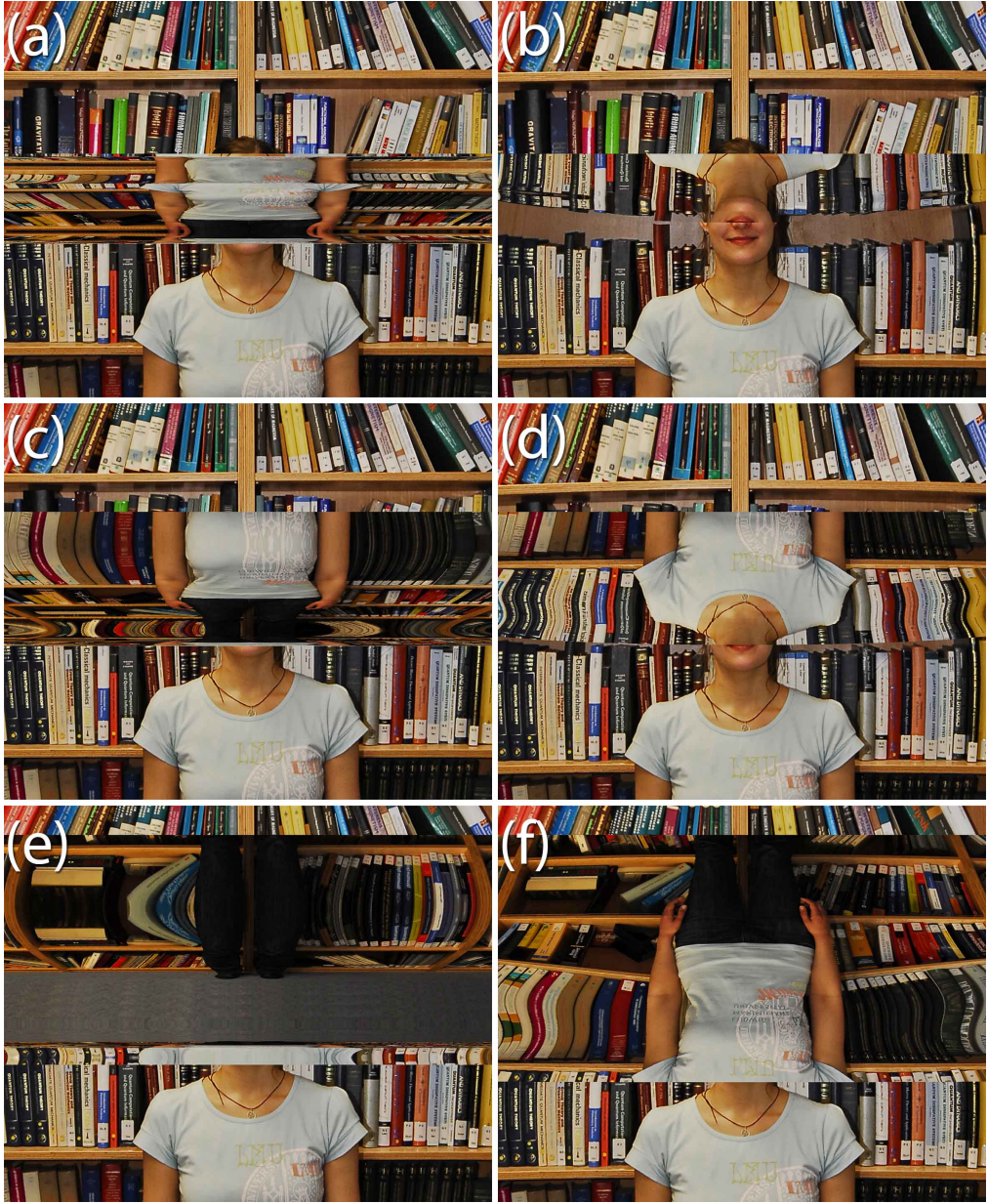

Fig. 9. The inhomogeneous locally isotropic double-Gaussian unidirectional free-space cloak for views away from the transformation axis. The mirror corrugation and its cloaking are shown for rotations of (a), (b) 5 degrees, (c), (d) 10 degrees, and (e), (f) 20 degrees. It is seen that the double-Gaussian unidirectional cloak behaves similarly to the Gaussian carpet cloak [15,16] for views away from the transformation axis. Fresnel-reflection coefficients for the above settings are 0.2% or less.

Design (iv) only exploits locally isotropic refractive indices, some of which are below unity. However, as done previously, one can *reference* this structure to unity *via* dividing the index distribution by its minimum value [9,16,18], thus achieving a graded-index minimum refractive index of unity. The free-space cloaking performance of (iv) in terms of range of viewing angles is not as good as (i)-(iii), and this is mainly due to the fact that to achieve a comparable cloaked volume, (iv) requires a cloaked region that is more spread along the transformation axis than in the case of (i)-(iii), thus leading to wider apparent mirrors upon views away from the transformation axis. In terms of performance in three dimensions while viewing along the transformation axis (Figs. 2, 4, 6, and 8), we see that (iii) is a perfect ray and wave cloak. However, its birefringent versions (i) and (ii) show a deterioration in cloaking performance away from the center vertical plane, although design (ii) still exhibits great 3D cloaking behavior for horizontally polarized light. (iv) shows amazingly good performance in three dimensions as well, making it a formidable experimentally feasible three-dimensional unidirectional free-space invisibility cloak. In miniature form, polarization insensitive three-dimensional structures like (iv) could for instance be fabricated in polymer form by direct-laser-writing optical laser lithography [11], combined with stimulated-emission depletion even at visible operation frequencies [12].

The following Table aims at helping the reader comparing the *photorealistic renderings* shown in this paper with all of those of our and other group's previous publications on three-dimensional transformation-optics architectures, including sceneries involving negative-index materials, which can be seen as resulting from a piecewise linear one-dimensional coordinate transformation [8].

| Type | Remarks | Polarization | Directional | Ref. |
|---|---|---|---|---|
| 3D carpet cloak | Quasi-conformal map, graded index | Independent | Omni | [18,19] |
| 3D carpet cloak | Conformal map, graded index | Independent | Omni | [16] |
| 3D carpet cloak | Periodic grating cloak, graded index | Independent | Omni | [16] |
| Free-space cloak | Impedance matched, cylindrical | Independent | Omni | [6,20] |
| Free-space cloak | Uniaxial birefringent dielectric, cylindrical | Dependent | Omni | [17] |
| Free-space cloak | Piecewise homogeneous, birefringent magnetic, cuboidal | Dependent | Uni | This work |
| Free-space cloak | Conformal map, graded index, cuboidal | Independent | Uni | This work |
| 360-degree sphere | "Invisible sphere", regularized | Dependent | Omni | [21] |
| 360-degree sphere | "Invisible sphere", graded index | Independent | Omni | [6] |
| 90-degree sphere | Graded index | Independent | Omni | [22] |
| 3D negative-index materials | Piecewise homogeneous | Independent | Omni | [21,23] |

**Acknowledgements**

We thank Ralph Wehrspohn (Universität Halle) for mentioning possible applications of invisibility cloaks for solar-cell applications to one of us (M.W.). We thank our model, Tanja Rosentreter (LMU München), and the photographer, Vincent Sprenger (TU München), for help with the photographs used as input for the photorealistic ray-tracing calculations. We thank Maximilian Papp (TU München) for his artistic help with Fig. 1. J.C.H. is grateful to Lode Pollet (LMU München) for fruitful discussions. We acknowledge the support of the Arnold Sommerfeld Center (LMU München), which allowed us to use their computer facilities for our rather CPU-time-consuming numerical ray-tracing calculations. J.C.H. acknowledges financial support from the Deutsche Forschungsgemeinschaft (DFG) through FOR801 and by the Excellence Cluster "Nanosystems Initiative Munich (NIM)". M.W. acknowledges support by the DFG, the State of Baden-Württemberg, and the Karlsruhe Institute of Technology (KIT) through the DFG Center for Functional Nanostructures (CFN) within subproject A1.5.